\documentclass[twocolumn,amssymb,amsmath,showpacs,prd]{revtex4}
\begin{document} 


\title{Radiation reaction in quantum field theory}


\author{Atsushi Higuchi}
\email{ah28@york.ac.uk}
\affiliation{
Department of Mathematics, University of York,\\ 
Heslington, York YO10 5DD, United Kingdom}
\date{March 30, 2004}


\begin{abstract}
We investigate radiation-reaction effects
for a charged scalar particle accelerated
by an external potential realized as a space-dependent mass term
in quantum electrodynamics.  
In particular, we calculate the position
shift of the final-state wave packet of the 
charged particle due to radiation
at lowest order in the fine structure constant $\alpha$ and in the
small $\hbar$ approximation.  
We show that it disagrees with the 
result obtained using the 
Lorentz-Dirac formula for the radiation-reaction force, and that
it agrees with the classical theory if one assumes
that the particle loses its
energy to radiation at each moment of time according 
to the Larmor formula in the
static frame of the potential.  However, the discrepancy is much 
smaller than the Compton wavelength of the particle. 
We also point out that the electromagnetic correction
to the potential has no classical limit. ({\bf Correction}. 
Surface terms were erroneously discarded to arrive at Eq.\ 
(\ref{intermed}).  By correcting this error we find that the
position shift according to the Lorentz-Dirac theory obtained from
Eq.\ (\ref{preDelz}) is reproduced by quantum field theory
in the $\hbar\to 0$ limit.  We also find that the small $V(z)$ 
approximation
is unnecessary for this agreement. See Sec.\ \ref{correction}.)
\end{abstract} 
\pacs{03.65.-w, 12.20.-m}

\maketitle


\section{Introduction}
\label{intro}

Although classical electrodynamics is a well established theory,
there has been some controversy over the nature of the reaction of a 
point charge to its own radiation.  The
standard equation for the radiation-reaction force, 
the Lorentz-Dirac 
equation\ \cite{Lorentz,Dirac},
admits runaway solutions which describe the charged particle 
accelerating on its own and reaching the speed of light in a very 
short time. (See, e.g., Refs.\ \cite{Jackson,Rohrlich} 
for a review of the Lorentz-Dirac theory.)
Although it has been shown that there is a unique solution 
that does not exhibit the runaway behaviour
for any bounded time-dependent force acting only for a finite interval 
of time\ \cite{Plass} (see also Refs.\ \cite{Rohr,Jackson}), 
this solution violates causality.  
It is generally believed that these problems will
be absent once the finite-size effect is properly taken into account.  
For example, an analysis of a charged sphere has led 
Yaghjian\ \cite{Yagh} to the conclusion that the radiation-reaction 
4-force is modified at the onset of acceleration in such a way that
there is no causality violation in the unique non-runaway solution.
There is also an approach based on 
``reduction of order"\ \cite{Landau} which excludes runaway 
solutions without introducing causality violation.  In this method
the radiation-reaction force is re-expressed in terms 
of the external force.  This approach has recently
been advocated by several authors\ \cite{FlanWald,Poisson}.

Thus, it may be fair to say that 
the aspects of the Lorentz-Dirac theory that were 
traditionally regarded as  
problematic have been clarified.
However, there are other counter-intuitive aspects in the 
Lorentz-Dirac theory which are not discussed very often.  These 
aspects persist in reformulations
of the Lorentz-Dirac equation mentioned before.  For example, the
radiation-reaction force vanishes if the acceleration of the charge is
constant.  Although 
this is not so serious a problem as existence of runaway solutions
or causality violation, it is certainly counter-intuitive since the 
charge radiates energy away continuously while being accelerated.
A related fact is that
the radiation-reaction force makes the kinetic energy
of the charge increase if the acceleration of the charge increases 
in time, as we will see in the next section, despite the fact
that the charge is radiating in the meantime.

In view of these counter-intuitive 
features of the Lorentz-Dirac theory, it is natural to ask
whether or not this theory reproduces the classical limit for a 
point charge in quantum electrodynamics
(QED). (Moniz and
Sharp have studied QED in this context and concluded that
the Lorentz-Dirac theory is reproduced in the classical limit by 
studying the Heisenberg equations for an extended 
charge and taking the zero-size limit\ \cite{Sharp}.)
It is well known that the loopwise expansion in quantum field theory 
is an expansion in powers of $\hbar$\ \cite{Nambu,BB}.  
Therefore the classical limit $\hbar \to 0$ in
this sense is the tree approximation.  However, one should not use 
this limit in the problem at hand.  For example, this limit is taken 
with the mass parameter $mc/\hbar$ held fixed (see,
e.g., Ref.\ \cite{IZ}).  Thus, the mass $m$ also tends to zero in this 
limit.  In the radiation-reaction problem the mass
as well as the electric charge $e$ must be kept fixed in the
limit $\hbar \to 0$.
Therefore, the limit we would need to consider is 
certainly not the tree approximation and, in fact, does not exist 
as emphasized recently by Julia\ \cite{Julia}. 
This fact can readily be seen by noting that
the fine structure constant
$\alpha = e^2/4\pi \hbar c\approx 1/137$ 
is of order $\hbar^{-1}$.  Thus, 
perturbation theory would break 
down in the $\hbar \to 0$ limit with $e$ fixed.  For example,
the one-loop 
correction to the electron magnetic moment, $\alpha/2\pi$, would 
diverge despite the fact 
that it is a very small fixed number.

Although the limit $\hbar \to 0$ with $m$ and $e$ fixed would be 
divergent, one can still compare some quantities in QED with 
classical ones at order $\alpha$.  At this order in $\alpha$, 
physical quantities of interest are of finite order in $\hbar^{-1}$,
and there are some quantities of order $\hbar^0$,
which can be compared with the corresponding quantities in classical
electrodynamics.  We note, however, that these quantities are much
smaller in reality than 
some quantities which vanish in the limit $\hbar \to 0$.
The following example illustrates this paradoxical feature.  
Let $\omega$ be the frequency of an electromagnetic wave.  
Then the energy of a single photon $\hbar\omega$
tends to zero as $\hbar \to 0$, but
$\alpha\hbar \omega = e^2\omega/4\pi c$ remains finite.
Thus, the quantity $\alpha \hbar \omega$ is meaningful in classical
electrodynamics even though it is much smaller than $\hbar \omega$.

In this paper we examine the behaviour of a wave packet of a charged 
scalar
field undergoing acceleration due to an external potential realized
as a position dependent mass term to first order
in $\alpha$.  
In particular, we calculate the change in the position of the 
final-state wave packet, the position shift, 
as a result of radiation using the WKB approximation.
We demonstrate that
the $\hbar \to 0$ limit of this quantity
disagrees with the corresponding result in the Lorentz-Dirac theory.
Then we show
that it agrees instead with the result in classical electrodynamics
obtained by assuming that the kinetic energy of the particle
is lost to the radiation given by the
classical Larmor formula at each moment of
time in the rest frame of the potential.
We also point out that the electromagnetic correction to the potential
is of order $\hbar^{-1}$ and, therefore, cannot be calculated in 
classical electrodynamics.  
The rest of the paper is organized as follows.
In Sec.~\ref{classical} we review the Lorentz-Dirac theory and
calculate the position shift of
a charged particle due to radiation in the nonrelativistic
approximation in this theory.
In Sec.~\ref{amplitude} we calculate the transition amplitude for
a charged scalar particle emitting a photon using the WKB 
approximation.  This is used in
Sec.~\ref{position}
to calculate the position shift
in scalar quantum electrodynamics.
Here, we closely follow an unpublished paper\ \cite{Higuchi}.  Then
in Sec.~\ref{forward}
we examine the one-loop effect (with no emission of photons) 
which results in a correction to the external potential.
We summarize our results and make some concluding remarks 
in Sec.~\ref{conclusion}.  The calculations are performed using
conveniently chosen physical polarization vectors for photons, but 
they are shown to agree with the results obtained without choosing
physical polarizations in Appendix A.  
A formula used to compute the position
shift is justified in Appendix B.
Our metric signature is $+---$.


\section{The position shift in the Lorentz-Dirac theory}
\label{classical}

We first motivate the standard Lorentz-Dirac theory briefly.
We do not derive it
but describe how it naturally arises in classical electrodynamics. 
(See, e.g. Ref.\ \cite{Rohrlich} for a derivation.)
Suppose that a classical charged particle with charge $e$ and
4-velocity $u^\mu$ is 
accelerated by 
an external 4-force $F^\mu_{\rm ext}$. [The 4-velocity of a 
particle moving in the $z$-direction with speed $v$ 
is $(\gamma c, 0,0, \gamma v)$ with
$\gamma \equiv (1-v^2/c^2)^{-1/2}$. Note that $u^\mu u_\mu = c^2$.]  
The 4-velocity $u^\mu$ satisfies the following equation:
\begin{equation}
m\frac{du^\mu}{d\tau} = F^\mu_{\rm ext}\,,  \label{noreac}
\end{equation}
where $m$ is the rest mass of the particle and where $\tau$ is the 
proper time along the world line of the particle. 
The particle emits electromagnetic
radiation with 4-momentum $P^{\mu} = (E/c,{\bf P})$, where $E$ is 
the energy and ${\bf P}$ is the momentum, given by the relativistic 
generalization of the Larmor formula:
\begin{equation}
\frac{dP^{\mu}}{d\tau} = \frac{e^2}{6\pi c^5} a^2 u^{\mu}\,, 
\label{Larmor}
\end{equation}
where $a = \sqrt{-\dot{u}^{\mu}\dot{u}_\mu}$ (with 
$\dot{u}^\mu = d u^\mu/d\tau$) 
is the proper acceleration of the particle
(see, e.g., Ref.\ \cite{Jackson,Rohrlich}).
Since it is the particle that
radiates, one expects that its 4-momentum
should be reduced by the amount carried away by the radiation. 
It is natural to describe this effect in terms of
a 4-force.  Thus, one is led to modify the equation of motion of the
particle as
\begin{equation}
m \frac{du^\mu}{d\tau} = F_{\rm ext}^\mu + K^\mu\,,  \label{Newton}
\end{equation}
where $K^\mu$ is the radiation-reaction 4-force which represents the
back reaction of the radiation on the charged particle.
It may seem
that the 4-force $K^\mu$ should be equal to the negative of
the right-hand
side of Eq.\ (\ref{Larmor}) because of energy-momentum conservation. 
However, this cannot be the case because
this 4-force would not satisfy the condition $u_\mu K^\mu =0$, which
is a consequence of the equation $u_\mu \dot{u}^\mu = 0$.
Therefore, one is led to require only that
the total loss of 4-momentum of the particle be equal to the
total 4-momentum radiated away assuming that the acceleration occurs 
only for a finite time.  Thus, one requires that
\begin{equation}
K^\mu = - \frac{e^2}{6\pi c^5}a^2 u^\mu + \frac{d C^\mu}{d\tau}
\end{equation}
for some vector $C^\mu$
because the second term does not contribute to the total change in the
4-momentum, $\int_{-\infty}^{+\infty} K^\mu\,d\tau$.  By using the 
condition $u_\mu K^\mu = 0$ one arrives at the following equation:
\begin{equation}
u_{\mu}\frac{dC^\mu}{d\tau} = \frac{e^2}{6\pi c^3}a^2\,.
\end{equation} 
Noting that $a^2 = - \dot{u}_\mu \dot{u}^\mu = u_\mu \ddot{u}^\mu$, 
one finds
\begin{equation}
\frac{dC^\mu}{d\tau} = \frac{e^2}{6\pi c^3}\frac{d^2 u^\mu}{d\tau^2}
+ \frac{d\tilde{C}^\mu}{d\tau}\,,
\end{equation}
where $d\tilde{C}^\mu/d\tau$ is orthogonal to $u^\mu$.
Letting $\tilde{C}^\mu = 0$, one arrives at
\begin{equation} 
K^{\mu} = \frac{e^2}{6\pi c^3} \left( 
\frac{d^2 u^{\mu}}{d\tau^2} - \frac{1}{c^2}a^2 u^\mu\right)\,.  
\label{LD}
\end{equation}
This is known as the Abraham 4-vector\ \cite{Abraham}, and 
Eq.\ (\ref{Newton}) with this 4-force is the Lorentz-Dirac equation.

To illustrate counter-intuitive features of this equation let
us consider one-dimensional motion parametrized by the rapidity 
$\beta$ with $u^\mu = (u^0, u^1) = (c\cosh \beta, c\sinh \beta)$ and
$F^\mu_{\rm ext} = (\tilde{F}_{\rm ext}\sinh\beta,
\tilde{F}_{\rm ext}\cosh\beta)$.
Then Eq.\ (\ref{Newton}) reads
\begin{equation}
\frac{d \beta}{d\tau} = \frac{1}{mc}\tilde{F}_{\rm ext} +
\frac{e^2}{6\pi mc^3}\frac{d^2 \beta}{d\tau^2}\,.
\label{2D}
\end{equation}
The second term
represents the radiation-reaction force.  Notice that it vanishes if
the acceleration is constant and
is positive if the acceleration increases.  Thus,
the radiation-reaction force pushes the particle forward if the
acceleration increases even though the particle is radiating.  This is
rather counter-intuitive. 

Now, let us describe the classical counterpart of our model.
In this model a particle with mass
$m$ and charge $e$ moves in the $z$-direction from 
$z=-\infty$ to $z=+\infty$ and is accelerated by
a static potential $V(z)$ whose derivative is 
nonzero only in a finite interval.  
We define $V_{-\infty} \equiv V(-\infty)$ and let $V(+\infty) = 0$.
We work in the nonrelativistic approximation.  Thus, 
the energy is given by $\frac{1}{2}mv^2 + V(z)$,
where $v$ is the velocity of the particle.
We assume that the kinetic energy is much larger than the potential 
energy, and we retain only the terms up to
second order in $V(z)$. Note that the particle moves in the positive
$z$-direction forever under this assumption.
We assume also that 
at $t=0$ the particle has passed the region with $V'(z)\neq 0$
and that it moves at constant velocity for $t>0$.
Let us denote the position and velocity of
the particle without radiation
by $\tilde{z}$ and $\tilde{v}$,
respectively, and define
$\delta z \equiv z - \tilde{z}$ and $\delta v \equiv v - \tilde{v}$.
One has $\delta z|_{t=-\infty} =0$ and $\delta v|_{t=-\infty} = 0$ by
definition.  
We will calculate the position shift due to radiation
at $t=0$ denoted by $\delta z|_{t=0}$ using the Lorentz-Dirac
equation.  We emphasize 
that this position shift is measured
after the acceleration because there is no acceleration for $t> 0$ by 
assumption.

The rate of change in the energy of the particle according to 
the Lorentz-Dirac theory is
\begin{equation}
\frac{d\ }{dt}\left[ \frac{1}{2}mv^2 + V(z)\right]
= \frac{e^2}{6\pi c^3}\ddot{v}v\,,  \label{LDenergy}
\end{equation}
where the dot indicates differentiation with respect to time $t$.
By writing $\ddot{v}v$ 
as $d(\dot{v}v)/dt - \dot{v}^2$ and 
integrating with respect to $t$, we find
\begin{equation}
\frac{1}{2}mv^2 +V(z) = E +  \frac{e^2}{6\pi c^3}v\dot{v}
 - \frac{e^2}{6\pi c^3}\int_{-\infty}^t \dot{v}^2\,dt\,, \label{ABRA}
\end{equation}
where $E$ is the initial energy.  
By subtracting the equation 
$\frac{1}{2}m\tilde{v}^2 + V(\tilde{z}) = E$ 
from Eq.\ (\ref{ABRA}) we find
to lowest order in $e^2$
\begin{equation}
m\tilde{v}\delta v + V'(\tilde{z})\delta z 
= \frac{e^2}{6\pi c^3}\tilde{v}\frac{d\tilde{v}}{dt}
- \frac{e^2}{6\pi c^3}\int_{-\infty}^t \tilde{a}^2\, dt\,, 
\label{Approx}
\end{equation}
where $\tilde{a} = d\tilde{v}/dt$.
By substituting $V'(\tilde{z}) = - m\,d\tilde{v}/dt$ in 
Eq.\ (\ref{Approx}) and dividing by $\tilde{v}^2$, we obtain
\begin{equation}
m\frac{d\ }{dt}\left(\frac{\delta z}{\tilde{v}}\right)
= \frac{e^2}{6\pi \tilde{v}c^3}\frac{d\tilde{v}}{dt} 
- \frac{e^2}{6\pi \tilde{v}^2c^3} \int_{-\infty}^{t} \tilde{a}^2\,dt\,.
\label{preDelz}
\end{equation}
By integrating this formula from $t=-\infty$ to $t=0$ we have, 
to second order in $V(z)$, 
\begin{equation}
\delta z|_{t=0} = \frac{e^2}{6\pi mc^3} v_f \log \frac{v_f}{v_i}
- \frac{e^2}{6\pi p c^3}\,I\,, \label{Delz}
\end{equation}
where $v_i$ and $v_f$ are the initial and final velocities, 
respectively, without radiation, and where
\begin{equation}
I \equiv  \int_{-\infty}^0 \left\{\int_{-\infty}^{t}[\tilde{a}(t')]^2
dt'\right\} dt\,.
\end{equation}
We have replaced $m\tilde{v}$ by the final momentum $p=mv_f$
in the second term in Eq.\ (\ref{Delz})
because we retain the terms only up 
to second order in $V(z)$ 
(and consequently to second order in the acceleration $\tilde{a}$).
Integrating by parts and using the assumption that 
$\tilde{a}(t) = 0$ for $t > 0$, we find
\begin{equation}
I  = -\int_{-\infty}^{+\infty} t\,[\tilde{a}(t)]^2\,dt\,.
\end{equation}
Now, the power of radiation can be found from Eq.\ (\ref{Larmor})
as
\begin{equation}
P_r(t) = \frac{e^2}{6\pi c^3}[\tilde{a}(t)]^2  \label{power}
\end{equation}
in the nonrelativistic approximation. (This is the Larmor formula.)
Then the position shift can be expressed as
\begin{equation}
\delta z|_{t=0} = \frac{e^2}{6\pi mc^3}v_f\log \frac{v_f}{v_i}
+\frac{1}{p}\int_{-\infty}^{+\infty} t P_r(t)\,dt\,.  \label{posshif2}
\end{equation}

In terms of the 
Fourier transform $\hat{a}_p(\omega)$ of $\tilde{a}(t)$ defined by
\begin{equation}
\hat{a}_p(\omega) = \int_{-\infty}^{+\infty}dt\, 
\tilde{a}(t)e^{i\omega t}\,,
\label{Fourier}
\end{equation}
we have
\begin{equation}
I =  i \int_{-\infty}^{+\infty} \frac{d\omega}{2\pi}\,
\hat{a}_p(\omega)^*\,
\frac{d\ }{d\omega}\,\hat{a}_p(\omega)\,,
\end{equation}
where $A^*$ denotes the complex conjugate of $A$.  (The subscript
``$p$" in $\hat{a}_p(\omega)$ has been inserted to 
emphasize that this quantity
depends on the momentum of the particle.  
This notation will turn out to be useful later.) Thus, to second 
order in $V(z)$ we have
\begin{eqnarray}
\delta z|_{t=0} & = & \frac{e^2}{6\pi m c^3}v_f\log\frac{v_f}{v_i}
\nonumber \\
&& - \frac{ie^2}{6\pi pc^3}
\int_{-\infty}^{+\infty}\,\frac{d\omega}{2\pi}\,
\hat{a}_p(\omega)^*\,\frac{d\ }{d\omega}\,\hat{a}_p(\omega)\,.  
\label{Shift}
\end{eqnarray}
The first term is first-order in the potential because $v_f - v_i$ is.
It is rather puzzling because the position shift is an effect caused
by radiation which is second-order in $V(z)$.  On the other hand,
the second term is clearly second-order.  Note also that the first
term would be absent if $v_f = v_i$, i.e. if $V_{-\infty}=0$.
In Sec.~\ref{position} we will derive the position shift
$\delta z|_{t=0}$ in the WKB approximation
in scalar QED.  We will find that the first term in 
Eq.\ (\ref{Shift}) is absent in the classical limit even if
$v_f \neq v_i$.


\section{The transition amplitude}
\label{amplitude}

We consider a complex scalar field $\psi(t,{\bf x})$ 
which is coupled to the electromagnetic field $A_\mu(t,{\bf x})$
and accelerated by an external
potential $V(z)$ with the properties described in the previous section.
This model is given by the following Lagrangian density:
\begin{eqnarray}
{\cal L} & = & [\partial_\mu - i(e/\hbar c)A_\mu]
\psi^\dagger\cdot
[\partial^\mu + i(e/\hbar c)A^\mu]\psi \nonumber \\
&& - \frac{1}{\hbar^2}[ m^2 c^2 + 2mV(z)]\psi^\dagger
\psi
- \frac{1}{4}F_{\mu\nu}F^{\mu\nu}\,, \label{Lagd}
\end{eqnarray}
with $\partial_0 \equiv c^{-1}\partial_t$ and
$F_{\mu\nu} \equiv \partial_\mu A_\nu - \partial_\nu A_\mu$.
The mode functions $\Phi(t,{\bf x})$ 
for the free scalar field (with $e=0$) satisfy
\begin{equation}
\left[ \partial_\mu \partial^\mu + m^2c^2/\hbar^2 
+ 2mV(z)/\hbar^2\right]
\Phi(t,{\bf x}) = 0\,.   \label{free}
\end{equation}
The energy $p_0^{\rm cl}$ and the momentum 
${\bf p}^{\rm cl}$ of the corresponding classical
particle satisfy
\begin{equation}
-(p_0^{\rm cl}/c)^2 + ({\bf p}^{\rm cl})^2 + m^2c^2 + 2mV(z)=0\,.
\end{equation}
(Note that the time component of the 4-momentum 
$p^{\rm cl}_\mu$ is not $p^{\rm cl}_0$ but $p_0^{\rm cl}/c$.)
In the nonrelativistic approximation with 
$m^2c^2 \gg ({\bf p}^{\rm cl})^2,\,\,2m|V(z)|$, we have
\begin{equation}
p_0^{\rm cl} \approx mc^2 + ({\bf p}^{\rm cl})^2/2m + V(z)\,.
\end{equation}
Thus, the particle moves under the influence of the nonrelativistic 
potential $V(z)$, and the quantum model given by Eq.\ (\ref{Lagd})
corresponds to the classical
one analyzed in the previous section.
Later we will assume $m^2c^2 \gg ({\bf p}^{\rm cl})^2 \gg 2m|V(z)|$ 
--- this condition
implies that the particle is nonrelativistic but the kinetic energy 
is much larger than the potential energy --- but we do not
use this assumption for the moment.

The solutions of Eq.\ (\ref{free}) which are relevant here can be 
written as 
\begin{equation}
\Phi_{p,{\bf p}_\perp}(t,{\bf x}) = 
\phi_p(z)\exp\left[ \frac{i}{\hbar}(-p_0 t + i{\bf p}_\perp \cdot 
{\bf x}_\perp)\right]\,,
\label{state}
\end{equation}
where ${\bf x}_\perp = (x,y)$, ${\bf p}_\perp = (p^x,p^y)$ and
\begin{equation}
p_0/c = \sqrt{m^2c^2 + p^2 + {\bf p}_\perp^2} \label{energy}
\end{equation}
with $p > 0$.
Thus, the quantity $p$ is the $z$-component of 
the momentum in the region with large and positive $z$ for which
the potential $V(z)$ vanishes.  The function $\phi_p(z)$ satisfies 
\begin{equation}
\left[ -\frac{\hbar^2}{2m} \frac{d^2\ }{dz^2} + V(z) \right]\phi_p(z)
= \frac{p^2}{2m}\phi_p(z)\,.  \label{non-rel}
\end{equation}
This function can be given in the WKB approximation (see, e.g., 
Ref.\ \cite{Schiff}) as
\begin{equation}
\phi_p(z) = \sqrt{\frac{p}{\kappa_p(z)}}\exp
\left[ \frac{i}{\hbar} \int_0^z \kappa_p(z')dz' + g(z)\right]
\label{wavefun}
\end{equation}
with 
\begin{equation}
\kappa_p(z) \equiv [p^2 - 2mV(z)]^{1/2}\,,  \label{kappa}
\end{equation}
where $g(z)$ is the correction
term of order $\hbar$.  We let $g(z)=0$
to work to lowest nontrivial
order in $\hbar$.
The wave function (\ref{wavefun}) is normalized so that
\begin{equation}
\int_{-\infty}^{+\infty}dz\,\phi_{p'}(z)^*\phi_p(z)
= 2\pi\hbar\delta(p-p')\,.
\end{equation}
Strictly speaking, this formula is not correct because the WKB 
approximation is not valid for some values of $p$.  
However, since we use only the modes
for which the WKB approximation is valid, the final result will not be 
affected even if we formally use this approximation for all range of 
$p$, as we do here.
Then, the field $\psi$ can be expanded using these modes as
\begin{equation}
\psi(x)  =  \hbar c \int \frac{d^3{\bf p}}{2p_0(2\pi\hbar)^3}
\left[ A({\bf p})\Phi_{\bf p}(x) + B^\dagger({\bf p})
\Phi_{\bf p}(x)^*\right]
\,,
\end{equation}
where $x\equiv (t,{\bf x})$, ${\bf p}\equiv (p,{\bf p}_\perp)$. 
The modes $\Phi_{\bf p}(t,{\bf x})$ are orthonormal:
\begin{equation}
i\hbar \int d^3{\bf x}\,\Phi_{\bf p}(t,{\bf x})^* 
\stackrel{\leftrightarrow}
{\partial_t} \Phi_{{\bf p}'}(t,{\bf x}) = 2p_0(2\pi\hbar)^3
\delta^3({\bf p}-{\bf p}')\,, \label{orthon}
\end{equation}
where 
$\stackrel{\leftrightarrow}{\partial_t}\, \equiv\, 
\stackrel{\rightarrow}{\partial_t} - 
\stackrel{\leftarrow}{\partial_t}$.
By imposing the usual
canonical commutation relations on $\psi(t,{\bf x})$,
\begin{subequations}
\begin{eqnarray}
\left[ \psi(t,{\bf x}), \partial_t \psi^\dagger(t,{\bf x}')\right]
& = & i\hbar c^2 \delta^3({\bf x}-{\bf x}')\,,\\
\left[ \psi^\dagger(t,{\bf x}), \partial_t \psi(t,{\bf x}')\right]
& = & i\hbar c^2 \delta^3({\bf x}-{\bf x}')\,, 
\end{eqnarray}
\end{subequations}
with all other equal-time commutators vanishing,
one finds
\begin{subequations}
\begin{eqnarray}
\left[ A({\bf p}), A^\dagger({\bf p}')\right] & = &
 2p_0(2\pi\hbar)^3
\delta^3({\bf p}-{\bf p}')\,,\\
\left[ B({\bf p}), B^\dagger({\bf p}')\right]
& =  & 2p_0(2\pi\hbar)^3
\delta^3({\bf p}-{\bf p}')\,.
\end{eqnarray}
\end{subequations}
All other commutators of the creation and annihilation operators 
vanish. The electromagnetic field in the Feynman gauge 
can be expanded in the usual manner as
\begin{equation}
A_\mu(t,{\bf x})  =  c \int \frac{d^3{\bf k}}{2\omega(2\pi)^3}
\left[ b_\mu ({\bf k})e^{-i\omega t + i{\bf k}\cdot {\bf x}}
 + {\rm h.c.}\right]\,,
\end{equation}
where $\omega = c\|{\bf k}\|$. 
The operators $b_\mu({\bf k})$ satisfy 
\begin{subequations}
\begin{eqnarray}
\left[ b_\mu ({\bf k}),b_\nu ({\bf k}')\right] & = & 0\,,\\ 
\left[ b_\mu({\bf k}),b_\nu^\dagger({\bf k}')\right] & = & 
- g_{\mu\nu} 2\hbar \omega (2\pi)^3\delta^3({\bf k}-{\bf k}')\,,
\end{eqnarray}
\end{subequations}
where $g_{\mu\nu}$ is the metric of the Minkowski
spacetime.
Notice that we have used the wave number ${\bf k}$ 
instead of the momentum to label the modes.  This is more convenient 
because the frequency and wave number of the emitted photon 
are classically well 
defined, the former being related to how rapidly the acceleration 
changes.  Thus, the momentum of the photon is of order $\hbar$.

We perform our calculations with physical polarization vectors
$\epsilon^{(j)\mu}({\bf k})$, $j=1,2$,
of the photon with wave number ${\bf k}$, 
which are real and with vanishing time components.  We choose them so
that $\epsilon^{(2)z}({\bf k}) = 0$.  It is shown 
in Appendix A that the calculations using the Fock space with 
indefinite metric in the Feynman gauge yield the same results. 
We define the transition amplitude from the initial state consisting
of a charged particle with momentum 
${\bf p} \equiv (p,{\bf p}_\perp)$ to the final state
consisting of a charged particle with momentum 
${\bf P} \equiv (P,{\bf P}_\perp)$ 
and a photon with wave number ${\bf k}$ and 
polarization vector $\epsilon^{(j)\mu}({\bf k})$ by
\begin{eqnarray}
&& {\cal A}(j,{\bf k},{\bf P},{\bf p}) \nonumber \\
&& \equiv
\frac{1}{\hbar}\int d^4 x \langle 0|
b^{(j)}({\bf k})A({\bf P})
{\cal L}_I(x)A^\dagger({\bf p})|0\rangle\,, \nonumber \\
\end{eqnarray}
where
\begin{equation}
b^{(j)}({\bf k}) \equiv \epsilon^{(j)\mu}({\bf k})b_\mu({\bf k})\,,
\end{equation}
with the following interaction Lagrangian density: 
\begin{equation}
{\cal L}_I = -\frac{ie}{\hbar c}
A^\mu\left( \psi^\dagger\partial_\mu \psi
-\partial_\mu\psi^\dagger\cdot \psi\right)\,.
\end{equation}
We only need the transition amplitude with the initial state
satisfying $p \gg \|{\bf p}_\perp\|$ because we will use 
a wave-packet state with this condition satisfied.
By a straightforward calculation we find
\begin{eqnarray}
{\cal A}(j,{\bf k},{\bf P},{\bf p})
& = & ec^2\int dz\left\{
({\bf p}_\perp + {\bf P}_\perp)\cdot 
\mbox{\boldmath $\epsilon$}^{(j)}_\perp({\bf k})
\phi_P(z)^*\,\phi_p(z) \right. \nonumber \\
&& \left. - i\hbar\epsilon^{(j)z}({\bf k})  
\phi_P(z)^*\stackrel{\leftrightarrow}{\partial}_z
\phi_p(z)\right\}e^{-ik_z z}\nonumber \\
&& 
\ \ \ 
\times (2\pi\hbar)^3 \delta^2({\bf P}_\perp
+\hbar{\bf k}_\perp -{\bf p}_\perp) \nonumber \\
&& \ \ \ \times\delta(P_0+\hbar \omega -p_0)  \label{first}
\end{eqnarray}
with ${\bf k}_\perp \equiv (k^x, k^y)$,
$\mbox{\boldmath $\epsilon$}^{(j)}_\perp 
\equiv (\epsilon^{(j)x}, \epsilon^{(j)y})$ and
$\stackrel{\leftrightarrow}{\partial}_z \equiv
\stackrel{\rightarrow}{\partial}_z - 
\stackrel{\leftarrow}{\partial}_z$.
We will calculate this transition amplitude to leading order in
$\hbar$.  Since we know on physical grounds
that the energy and momentum of the photon emitted are of order
$\hbar$ and, therefore, much smaller than those of the charged 
particle, we can assume that $P \gg \|{\bf P}_\perp\|$ for the final 
state because of the assumption that $p \gg \|{\bf p}_\perp\|$ for 
the initial state.  Thus, we neglect the term proportional to 
$({\bf p}_\perp + {\bf P}_\perp)\cdot 
\mbox{\boldmath $\epsilon$}^{(j)}_\perp({\bf k})$ 
in Eq.\ (\ref{first}).
Then, to lowest nontrivial order in $\hbar$ we find
\begin{eqnarray}
{\cal A}(j,{\bf k},{\bf P},{\bf p})
& = & ec^2\epsilon^{(j)z}({\bf k})G(k_z,P,p)\nonumber \\
&& \times (2\pi\hbar)^3 \delta^2({\bf P}_\perp+\hbar 
{\bf k}_\perp -{\bf p}_\perp) \nonumber \\
&& \times \delta(P_0+\hbar \omega -p_0)\,,   \label{AG}
\end{eqnarray}
where
\begin{equation}
G(k_z,P,p) \equiv
\int dz\, \phi_P(z)^*\left[\kappa_p(z) + \kappa_P(z)\right]
\phi_p(z)e^{-ik_z z}\,. \label{Gkz}
\end{equation}

Next we evaluate the integral $G(k_z,P,p)$ to lowest nontrivial order
in $\hbar$.
We keep only the terms which are 
first-order in $V(z)$ because
$G(k_z,P,p)$ will be squared in the position shift
for which we need only the second-order terms. 
First we note that to first order in $V(z)$ we
can write
\begin{equation}
G(k_z,P,p)
= 2\sqrt{Pp}\int dz\exp\left\{i
\int_0^z K(z')\,dz'\right\}\,,  \label{gkz}
\end{equation}
where
\begin{equation}
K(z) \equiv
\left[ \kappa_p(z) -\kappa_P(z)\right]/\hbar - k_z\,. \label{defK}
\end{equation}
This follows from the fact that
$[\kappa_p(z)/\kappa_P(z)]^{1/2} + [\kappa_P(z)/\kappa_p(z)]^{1/2}-2$
is second-order in $V(z)$.
Now, for any function $g(z)$, one has
\begin{equation}
\int_{a}^{b}dz\,e^{ig(z)}
= -i\left[\frac{e^{ig(z)}}{g'(z)}\right]_{a}^{b}
-i \int_{a}^{b}dz\,\frac{g''(z)}{\left[g'(z)\right]^2}e^{ig(z)}\,.
\label{gzform}
\end{equation}
We use this formula with $g(z) = \int_0^z K(z)\,dz + i\epsilon |z|$,
where the term $i\epsilon |z|$ with $\epsilon > 0$
has been inserted to regularize the
integral.  The boundary terms vanish because
one cannot have $K(z)=0$ as well as
\begin{subequations}
\begin{eqnarray}
p_0 & = & P_0 + \hbar \omega\,, \label{encon} \\
{\bf p}_\perp & = & {\bf P}_\perp + \hbar {\bf k}_\perp\,,
\label{momcon} 
\end{eqnarray}
\label{enmom}
\end{subequations}
which are enforced by the $\delta$-functions in Eq.\ (\ref{AG}),
without violating the mass-shell conditions.
Hence we have
\begin{eqnarray}
G(k_z,P,p) & = & 
-2i\sqrt{Pp}\int_{-\infty}^{+\infty} dz\,
\frac{K'(z)}{\left[K(z)\right]^2}\nonumber \\
&& \ \ \ \ \ \times 
\exp\left[ i\int_0^z K(z')\,dz'\right]\,. \label{Gkz1}
\end{eqnarray}

Let us find a small $\hbar$ approximation for $K(z)$.
By squaring the energy-conservation equation (\ref{encon}), we find
\begin{equation}
p^2 + {\bf p}_\perp^2 
\approx P^2 + {\bf P}_\perp^2 + 2m\hbar \omega  \label{Peq}
\end{equation}
in the nonrelativistic approximation and to first order in $\hbar$.
By using the inequality 
\begin{equation}
|{\bf p}_\perp^2 - {\bf P}_\perp^2| 
\leq \|{\bf p}_\perp - {\bf P}_\perp\|
 \|{\bf p}_\perp + {\bf P}_\perp\|
\end{equation}
and Eq.\ (\ref{momcon}), we find
\begin{equation}
|{\bf p}_\perp^2 -{\bf P}_\perp^2| 
\leq \hbar \omega/c\cdot\|{\bf p}_\perp + {\bf P}_\perp\|
\ll m\hbar \omega\,.
\end{equation}
Using this in Eq.\ (\ref{Peq}) we have
\begin{equation}
p^2 - P^2 \approx 2m\hbar \omega\,.  \label{good}
\end{equation} 
Hence we obtain the following approximate formula:
\begin{equation}
\kappa_p(z) - \kappa_P(z) \approx 
\frac{\partial\kappa_P(z)}{\partial P^2}
(p^2 - P^2) \approx \frac{\hbar \omega}{v_p(z)}\,,
\end{equation}
where 
\begin{equation}
v_p(z) \equiv \kappa_p(z)/m = \sqrt{p^2-2mV(z)}/m
\end{equation}
is the speed of the corresponding classical particle. 
Hence,
\begin{equation}
K(z) \approx \frac{\omega}{v_p(z)} - k_z\,. \label{Kappr}
\end{equation}
Furthermore,  since the charged particle
is nonrelativistic, we have
$\omega/v_p(z) = c\|{\bf k}\|/v_p(z) \gg k_z$.
Thus,
\begin{equation}
K(z) \approx \frac{\omega}{v_p(z)}\,. \label{Kappr2}
\end{equation}

By substituting Eq.\ (\ref{Kappr2})
in Eq.\ (\ref{Gkz1}) 
and letting $\sqrt{Pp}\approx p$, we obtain
\begin{eqnarray}
&& G(k_z,P,p)\nonumber \\
&& \approx 
\frac{2ip}{\omega}
\int_{-\infty}^{+\infty} dz
\,v_p'(z)
\exp\left\{ i\omega\int_0^z [v_p(z')]^{-1}\,dz'\right\}\,.\nonumber \\
&&  \label{Gappr}
\end{eqnarray}
Let us define the function $T(z)$ by
\begin{equation}
T \equiv \int_0^z [v_p(z')]^{-1}dz' + T_0\,,  \label{defT}
\end{equation} 
where $T_0$ is a constant.
Then by using
\begin{equation}
dz\, v_p'(z) = dz\,\frac{dv_p}{dz} = dT\,\frac{dv_p}{dT}
\end{equation}
and noting that $T\to \pm \infty$ 
as $z\to \pm \infty$ because $v_p(z)$ is always positive, we obtain
\begin{equation}
G(k_z,P,p) =
\frac{2ip}{\omega}
\int_{-\infty}^{+\infty} dT\,\frac{dv_p}{dT} e^{i\omega (T-T_0)}\,.
\label{intermed}
\end{equation}
Now, Eq.\ (\ref{defT}) implies that $dz/dT = v_p$.  Thus, the
quantity $T$ can be identified with the time coordinate of the
classical particle if the constant $T_0$ is chosen appropriately.
Recall that we need to evaluate 
$G(k_z,P,p)$ only to first order in $V(z)$. 
Since the acceleration $dv_p/dT$ is first-order
in $V(z)$, we need to find $T_0$ only to zeroth order in $V(z)$,
i.e. we may neglect the potential.  Since
$T = T_0$ for $z=0$, the constant $T_0$ is the time when the particle
is at $z=0$.  Letting the particle be at $z=z_0$ for $T=0$, we have
$z_0/T_0 = - v_p = -p/m$.  Hence $T_0 = -mz_0/p$.  By substituting this
in Eq.\ (\ref{intermed}) we find
\begin{equation}
G(k_z,P,p) = 
\frac{2ip}{\omega}\hat{a}_p(\omega)\exp(im\omega z_0/P)\,, \label{Gpk}
\end{equation}
where the Fourier transform of the acceleration, $\hat{a}_p(\omega)$,
is defined by Eq.\ (\ref{Fourier}) [with the identification
$dv_p/dT = \tilde{a}(t)$].
One may replace $p$ by $P$ and vice versa in Eq.\ (\ref{Gpk})
because the difference $p-P$ is of order $\hbar$.

It is instructive to compare this result with
the transition amplitude for radiation
due to a classical current with the 
following interaction Lagrangian density:
\begin{equation}
{\cal L}_{I\,{\rm cl}} = (e/c)A_\mu(t,{\bf x}) j^\mu(t,{\bf x})\,.
\end{equation}
Here, the current density $j^\mu(x)$ is given by
\begin{subequations}
\begin{eqnarray}
j^0(t,{\bf x}) & = & c\delta^2({\bf x}_\perp)\delta[z - z(t)]\,,\\
j^z(t,{\bf x}) & = & v(t)\delta^2({\bf x}_\perp)\delta[z-z(t)]
\end{eqnarray}
\end{subequations}
with $j^x(t,{\bf x}) = j^y(t,{\bf x}) = 0$,
where $z(t)$ is the position of the charged particle at time $t$ and 
where $v(t) = z'(t)$ is its velocity in the $z$-direction. 
The transition amplitude from the zero-photon state
$|0\rangle_\gamma$ to a one-photon state
$b^{(j)\dagger}({\bf k})|0\rangle_\gamma$ is
\begin{eqnarray}
{\cal A}_{\rm cl}(j,{\bf k}) & \equiv & 
\frac{1}{\hbar}\int d^4 x\, {}_\gamma\langle 0|
b^{(j)}({\bf k}) 
{\cal L}_{I\,{\rm cl}}(x)|0\rangle_\gamma \nonumber \\
& = & e\epsilon^{(j)z}({\bf k}) 
\int_{-\infty}^{+\infty}dt\, v(t)\exp\left[
i\omega t - ik_z z(t)\right]\,. \nonumber \\
\end{eqnarray}
By integrating by parts with a damping factor
$e^{-\epsilon|t|}$ and dropping the surface terms because
$\omega - k_zv(z)\neq 0$ for all $z$, we obtain
\begin{equation}
{\cal A}_{\rm cl}(j,{\bf k}) 
=  ie\epsilon^{(j)z}({\bf k})\omega\int_{-\infty}^{+\infty}dt
\frac{v'(t)\exp\left[ i\omega t - ik_z z(t)\right]}
{\left[\omega - k_zv(t)\right]^2}\,.
\end{equation}
Since $\omega = c\|{\bf k}\| \gg k_z v(t)$ and
$\omega t \gg k_z z(t)$, we find
\begin{eqnarray}
{\cal A}_{\rm cl}(j,{\bf k}) &\approx & 
\frac{ie\epsilon^{(j)z}({\bf k})}{\omega}
\int_{-\infty}^{+\infty}dt\,v'(t)e^{i\omega t}\nonumber \\
& = & \frac{ie\epsilon^{(j)z}({\bf k})}{\omega}\hat{a}_p(\omega)\,.
\end{eqnarray}
Notice the similarity of this expression with Eq.\ (\ref{Gpk}).
Our result obtained in the small $\hbar$ approximation
is closely related to the emission process by the corresponding
classical particle.

\section{The position shift in quantum electrodynamics}
\label{position}

Now, let us consider an initial wave-packet state given by
\begin{equation}
|i\rangle = \int\frac{d^3{\bf p}}{\sqrt{2p_0}(2\pi\hbar)^3}
f({\bf p})A^\dagger({\bf p})|0\rangle\,.
\label{wave_packet}
\end{equation}
The normalization condition $\langle i\,|\,i\rangle =1$ leads to
\begin{equation}
\int \frac{d^3{\bf p}}{(2\pi\hbar)^3}|f({\bf p})|^2 = 1\,.
\label{normall}
\end{equation}
The function $f({\bf p})$ will be identified later 
as the nonrelativistic
one-particle wave function in the momentum representation.
We choose the following form:
\begin{equation}
f({\bf p}) = |f({\bf p})| e^{-ipz_0/\hbar} \label{concrete}
\end{equation}
with
\begin{equation}
|f({\bf p})| = N 
\exp\left[ -\frac{(p-\overline{p})^2 
+ \|{\bf p}_\perp\|^2}{2(\Delta p)^2} \right]\,,
\end{equation}
where $N$ is the normalization constant.
Here, $\Delta p$ and $\overline{p}\,(>0)$ 
are the width of the wave packet in
the momentum space and the
average momentum [in the region with $V(z)=0$], respectively.
The constant $z_0$ will turn out to be 
the average $z$-coordinate at $t=0$.
We now assume that 
$m^2c^2\gg \overline{p}^2 \gg m|V(z)|$.  We also assume that
$\Delta p \ll \overline{p}$.
The transverse momentum ${\bf p}_\perp$ 
is much smaller than $\overline{p}$ under this assumption. 
This wave packet describes a nonrelativistic charged particle 
moving in the positive $z$-direction with its 
kinetic energy much larger 
than $|V(z)|$.  The initial state $|i\rangle$ given by 
Eq.\ (\ref{wave_packet}) leads to the following one-photon final 
state according to the standard time-dependent perturbation theory 
in the interaction picture: 
\begin{eqnarray}
|f,{\rm 1\gamma}\rangle 
& = & i\sum_{j=1}^2
\int\frac{d^3{\bf k}}{2\hbar \omega(2\pi)^3}
\int\frac{d^3{\bf P}}{2P_0(2\pi\hbar)^3} \nonumber \\
&& \times \int\frac{d^3{\bf p}}{\sqrt{2p_0}(2\pi\hbar)^3} 
\,f({\bf p})
{\cal A}(j,{\bf k},{\bf P},{\bf p})\nonumber \\
&& \times b^{(j)\dagger}({\bf k}) A^\dagger({\bf P})|0\rangle\,,
\label{final}
\end{eqnarray}
where $P_0/c \equiv \sqrt{m^2c^2 + P^2 + {\bf P}_\perp^2}$.
The amplitude ${\cal A}(j,{\bf k},{\bf P},{\bf p})$ is defined by
Eq.\ (\ref{AG}) with $G(k_z,P,p)$ given by Eq.\ (\ref{Gpk}).
By substituting Eq.\ (\ref{AG}) in Eq.\ (\ref{final}) 
and using, for any function $F({\bf p})$,
\begin{eqnarray}
&& 
\int\frac{d^3{\bf p}}{\sqrt{2p_0}}F({\bf p})
\delta^2({\bf p}_\perp + \hbar{\bf k}_\perp
- {\bf P}_\perp)\delta(P_0+\hbar\omega-p_0)\nonumber \\
&& = \frac{\sqrt{2p_0}}{2pc^2}F({\bf p})\,,
\end{eqnarray}
where ${\bf p}$ on the right-hand side
is expressed in terms of ${\bf P}$ and ${\bf k}$ through the
relations (\ref{enmom}),
and recalling that $\epsilon^{(2)z}({\bf k}) = 0$, we obtain
\begin{eqnarray}
|f,1\gamma\rangle & = & \frac{ie}{\hbar}
\int\frac{d^3{\bf k}}{2\omega(2\pi)^3}
\int\frac{d^3{\bf P}}{2P_0(2\pi\hbar)^3}
\frac{\sqrt{2p_0}}{2p}f({\bf p})
\nonumber \\
&& \times
\epsilon^{(1)z}({\bf k})G(k_z,P,p)
b^{(1)\dagger}({\bf k}) A^\dagger({\bf P})|0\rangle\,.
\label{fnal}
\end{eqnarray}
Now, from Eq. (\ref{good}) we have $p-P \approx m\hbar \omega/P$.  
By using this approximation in 
Eq.\ (\ref{concrete}) we find
\begin{equation}
f({\bf p})
\approx |f({\bf p})|\exp\left(-iPz_0/\hbar -im\omega z_0/P\right)\,.
\label{fapprox}
\end{equation}
By substituting this equation and Eq.\ (\ref{Gpk})
in Eq.\ (\ref{fnal}) we obtain
\begin{eqnarray}
|f,1\gamma\rangle & = & -\frac{e}{\hbar}
\int\frac{d^3{\bf k}}{2\omega^2(2\pi)^3}
\int\frac{d^3{\bf P}}{\sqrt{2P_0}(2\pi\hbar)^3}\nonumber \\
&& \times |f({\bf p})|\exp\left( -iPz_0/\hbar\right)
\epsilon^{(1)z}({\bf k})\hat{a}_p(\omega)\nonumber \\
&& \times b^{(1)\dagger}({\bf k}) A^\dagger({\bf P})|0\rangle\,,
\label{realfinal}
\end{eqnarray}
where
\begin{equation}
{\bf p} = (p,{\bf p}_\perp) = (P + m\hbar\omega/P,{\bf P}_\perp 
+ \hbar{\bf k}_\perp)\,. \label{PPP}
\end{equation}
We have approximated $p_0$ by $P_0$ on the right-hand side
of Eq.\ (\ref{realfinal}).

The emission probability can be obtained to leading order in $\hbar$
from Eq.\ (\ref{realfinal}) as
\begin{eqnarray}
{\cal P}_r & \equiv & \langle f,1\gamma\,|\,f,1\gamma\rangle
\nonumber \\
& = & \frac{e^2}{\hbar} 
\int\frac{d^3{\bf k}}{2\omega^3(2\pi)^3}
\int\frac{d^3{\bf P}}{(2\pi\hbar)^3} \nonumber \\
&& \ \ \ \ \ \times 
|f({\bf p})|^2 \left[\epsilon^{(1)z}({\bf k})\right]^2
|\hat{a}_p(\omega)|^2 \nonumber \\
& \approx & \frac{e^2}{\hbar} 
\int\frac{d^3{\bf k}}{2\omega^3(2\pi)^3}
\left[\epsilon^{(1)z}({\bf k})\right]^2
|\hat{a}_{\overline{p}}(\omega)|^2\,,
\end{eqnarray}
where we have used the fact that the function $|f({\bf p})|$ is
sharply peaked about $(p,{\bf p}_\perp)=(\overline{p},0)$.
Recalling that 
$d^3{\bf k} = d\Omega_{{\bf k}}\omega^2\,d\omega/c^3$, where
$d\Omega_{{\bf k}}$ is the solid-angle element in the ${\bf k}$ space,
and that the 
average of $|\epsilon^{(j)z}({\bf k})|^2$ for each value of $\omega$ 
is $2/3$, we find
\begin{equation}
{\cal P}_r
= \frac{e^2}{3\pi \hbar c^3}\int_0^\infty \frac{d\omega}{2\pi \omega}
|\hat{a}_{\overline{p}}(\omega)|^2\,.  \label{transprob}
\end{equation}
Notice that this probability is infrared divergent unless 
$\hat{a}_{\overline{p}}(0) = V_{-\infty}/{\overline{p}}=0$.
However, the expected energy of the photon is finite and is given by 
\begin{eqnarray}
E_r & = &
\frac{e^2}{3\pi 
\hbar c^3}\int_0^\infty \frac{d\omega}{2\pi \omega}\hbar\omega 
|\hat{a}_{\overline{p}}(\omega)|^2\nonumber \\
& = & \frac{e^2}{6\pi c^3}\int_{-\infty}^{+\infty}dt\,
\left[a_{\overline{p}}(t)\right]^2\,, \label{lar}
\end{eqnarray}
where $a_{\overline{p}}(t)$ is the acceleration of the corresponding
classical particle at time $t$.
This is nothing but the Larmor formula (\ref{power}).

For the classical current discussed at the end of the previous
section, the one-photon final state is
\begin{eqnarray}
|f,1\gamma\rangle_{\rm cl} & = & 
\frac{i}{\hbar}\int \frac{d^3{\bf k}}{2\omega (2\pi)^3}
{\cal A}_{\rm cl}(1,{\bf k})b^{(1)\dagger}({\bf k})|0\rangle_\gamma
\nonumber \\
& = & -\frac{e}{\hbar}\int\frac{d^3{\bf k}}{2\omega^2(2\pi)^3}
\epsilon^{(1)z}({\bf k})\hat{a}_p(\omega)b^{(1)\dagger}({\bf k})
|0\rangle_\gamma\,.\nonumber \\  
\end{eqnarray}
The emission probability calculated from this state agrees with
Eq.\ (\ref{transprob}).

Now, we derive the quantum expression of the position shift
to order $e^2$ in the small $\hbar$ approximation
and compare it with the result in the Lorentz-Dirac theory derived in
Sec.~\ref{classical}.  
We assume that the constant $z_0$ 
in Eq.\ (\ref{concrete}) is larger than any
$z$ with $V(z) \neq 0$ and that the wave packet is
far into the region with $V(z) = 0$ at $t=0$. 
First we note that the expected value 
of the $z$-coordinate at $t=0$ with the 
electromagnetic field turned off
is given by
\begin{equation}
\langle z \rangle_{t=0}^{\rm off} = \frac{i\hbar}{2}\int
\frac{d^3{\bf p}}{(2\pi\hbar)^3}
\left[f({\bf p})^*\frac{\partial f({\bf p}) }{\partial p} 
 - \frac{\partial f({\bf p})^*}{\partial p}
\,f({\bf p})\right]\,, \label{Peval}
\end{equation}
where $f({\bf p})$ is defined by Eq.\ (\ref{concrete}).
This formula can be
understood as the statement that the function $f({\bf p})$ is the 
momentum representation of the one-particle wave function.  
The derivation of Eq.\ (\ref{Peval}) is given in Appendix B.
Since
\begin{equation}
\frac{i\hbar}{2}\left[f({\bf p})^*
\frac{\partial f({\bf p}) }{\partial p} 
 - \frac{\partial f({\bf p})^*}{\partial p}
\,f({\bf p})\right] = z_0|f({\bf p})|^2\,, \label{z0}
\end{equation}
we find
\begin{equation}
\langle z \rangle_{t=0}^{\rm off} = z_0\,.
\end{equation}

Next we calculate the expected value of 
the $z$-coordinate at $t=0$ with the 
electromagnetic field turned on.  
At lowest nontrivial order in $e$,  the final
state can be given as
\begin{equation}
|f\rangle = |f,0\gamma\rangle + |f,1\gamma\rangle\,,
\end{equation}
where $|f,1\gamma\rangle$ is the one-photon final state given by 
Eq.\ (\ref{realfinal}) and where the zero-photon final state 
$|f,0\gamma\rangle$ can be written as
\begin{equation}
|f,0\gamma\rangle = \int \frac{d^3{\bf p}}{\sqrt{2p^0}(2\pi\hbar)^3}
[1+i{\cal F}({\bf p})]f({\bf p})A^\dagger({\bf p})|0\rangle\,.
\end{equation}
We will not explicitly evaluate the one-loop
forward-scattering amplitude ${\cal F}({\bf p})$, 
which is of order $e^2$.
The expected value of the $z$-coordinate at $t=0$ is the sum of the
zero-photon and one-photon contributions.
The zero-photon contribution, $\langle z\rangle_{t=0}^{(0)}$, can be
obtained by replacing 
$f({\bf p})$ by $[1+i{\cal F}({\bf p})]f({\bf p})$
in Eq.\ (\ref{Peval}).  The result is
\begin{equation}
\langle z\rangle_{t=0}^{(0)}
= \int \frac{d^3{\bf p}}{(2\pi\hbar)^3} |f_{\bf p}|^2
\left[
z_0(1-2{\rm Im}\,{\cal F})
- \hbar \frac{\partial\ }{\partial p}
{\rm Re}\,{\cal F}\right]
\,, \label{zero-ph}
\end{equation}
where we have written $f({\bf p}) \equiv f_{\bf p}$ and used
Eq.\ (\ref{z0}).
The contribution from the one-photon state,
$\langle z\rangle_{t=0}^{(1)}$, is similar 
except that one needs to trace over
the photon states. Recall that 
we have calculated $G(k_z,P,p)$ in Eq.\ (\ref{Gpk})
to lowest order in $\hbar$.  We define 
$\hat{a}_1({\bf k},{\bf p})$ 
as a quantity such that the expression for
$G(k_z,P,p)$ in that equation will be valid to next order in $\hbar$
if $\epsilon^{(1)z}({\bf k})\hat{a}_p(\omega)$ 
is replaced by it. (Introduction of
$\hat{a}_1({\bf k},{\bf p})$ is necessary
because $\langle z\rangle_{t=0}^{(0)}$ and 
$\langle z\rangle_{t=0}^{(1)}$ are of order $\hbar^{-1}$.)
Then, we have, to order $\hbar^0$,
\begin{eqnarray}
\langle z\rangle_{t=0}^{(1)} & = & 
\frac{ie^2}{2}
\int \frac{d^3{\bf k}}{2\omega^3(2\pi)^3}
\int \frac{d^3{\bf P}}{(2\pi\hbar)^3}\,
\nonumber \\
&& \times \left[ \left|\hat{a}_1({\bf k},{\bf p})\right|^2
\left( f_{\bf p}^*\frac{\partial f_{\bf p}}{\partial P}
- \frac{\partial f_{\bf p}^*}{\partial P}
f_{\bf p} \right) \right. \nonumber \\
&&  +|\epsilon^{(1)z}({\bf k})|^2 \left|f_{\bf p}\right|^2
\nonumber \\
&& \left. \times
\left( \hat{a}_p(\omega)^*\frac{\partial\hat{a}_p(\omega)}{\partial P}
- \frac{\partial\hat{a}_p(\omega)^*}{\partial P}\,
\hat{a}_p(\omega)\right)\right]\,,\nonumber \\
\label{one-ph}
\end{eqnarray}
where ${\bf p}$ is related to ${\bf P}$ by Eq.\ (\ref{PPP}).
The integration variables ${\bf P}$ can be changed to
${\bf p}$ if we change the derivative $\partial/\partial P$
to $\partial/\partial p$ at the same time.  Then by using
Eq.\ (\ref{z0}) again we find
\begin{eqnarray}
\langle z\rangle_{t=0}^{(1)} & = &
\int \frac{d^3{\bf p}|f_{\bf p}|^2}{(2\pi\hbar)^3}
\int \frac{d^3{\bf k}}{2\omega^3(2\pi)^3}\nonumber \\
&& \times \left[ \frac{e^2 z_0}{\hbar}\left|\hat{a}_1({\bf k},{\bf p})
\right|^2 \right. 
+\frac{ie^2}{2}|\epsilon^{(1)z}({\bf k})|^2 \nonumber \\
&& \left. \times 
\left( \hat{a}_p(\omega)^*\frac{\partial\hat{a}_p(\omega)}{\partial p}
-\frac{\partial\hat{a}_p(\omega)^*}{\partial p}\,
\hat{a}_p(\omega)\right)\right]\,,\nonumber \\
\end{eqnarray}
Unitarity implies that
$\langle f,0\gamma\,|\,f,0\gamma\rangle + 
\langle f,1\gamma\,|\,f,1\gamma\rangle  = 1$
at order $e^2$. 
This results in the following relation:
\begin{equation}
2\,{\rm Im}\,{\cal F} =
\frac{e^2}{\hbar} \int \frac{d^3{\bf k}}{2\omega^3(2\pi)^3}
|\hat{a}_1({\bf k},{\bf p})|^2\,. 
\label{ImF}
\end{equation}
Both sides are equal to the emission probability, 
which is given by Eq.\ (\ref{transprob}) at leading order in $\hbar$.
By adding $\langle z\rangle_{t=0}^{(0)}$ and 
$\langle z\rangle_{t=0}^{(1)}$ given by 
Eqs.\ (\ref{zero-ph}) and (\ref{one-ph}), respectively, using
Eq.\ (\ref{ImF}) and recalling that the function
$f_{\bf p}$ is sharply peaked about 
$(p,{\bf p}) = (\overline{p},0)$ we have
\begin{eqnarray}
\langle z\rangle_{t=0}
& \approx & z_0 -\left. \hbar
\frac{\partial\ }{\partial p}{\rm Re}\,
{\cal F}\right|_{p=\overline{p},{\bf p}_\perp=0}\nonumber \\
&& + \frac{ie^2}{2}\int
\frac{d^3{\bf k}|\epsilon^{(1)z}({\bf k})|^2}{2\omega^3(2\pi)^3}
\nonumber \\
&& \times \left[ \hat{a}_p(\omega)^*\frac{\partial\ }{\partial p}
\hat{a}_p(\omega) - \frac{\partial\ }{\partial p}\hat{a}_p(\omega)^*
\cdot\hat{a}_p(\omega)\right]_{p=\overline{p}}\,.\nonumber \\
\label{intmed}
\end{eqnarray}
Now we note that the classical acceleration 
depends on $p$ and $t$ through 
$z = pt/m+z_0$. [One can use the
free-particle approximation here because the last term 
in Eq.\ (\ref{intmed}), where $\hat{a}_p(\omega)$ appears,
is already second-order in
$V(z)$.]  Thus,
\begin{equation}
\left( t\,\frac{\partial\ }{\partial t} 
- p\,\frac{\partial\ }{\partial p}
\right)\,\int_{-\infty}^{+\infty}
\frac{d\omega}{2\pi} \hat{a}_p(\omega)e^{-i\omega t}
= 0\,. \label{vital}
\end{equation}
Then we find
\begin{equation}
\frac{\partial\ }{\partial p}\hat{a}_p(\omega)
= - \frac{1}{p}\hat{a}_p(\omega) 
- \frac{\omega}{p}\frac{\partial\ }{\partial \omega}
\hat{a}_p(\omega)\,.
\end{equation}
{}From this formula and the fact that
$\hat{a}_p(-\omega) = \hat{a}_p(\omega)^*$ [because the acceleration
is real], we obtain
\begin{eqnarray}
\langle z\rangle_{t=0} & \approx & z_0
-\left.\hbar\frac{\partial\ }{\partial p}
{\rm Re}\,{\cal F}\right|_{p=\overline{p},{\bf p}_\perp=0}\nonumber \\
&& - \frac{ie^2}{6\pi pc^3}
\int_{-\infty}^{+\infty}
\frac{d\omega}{2\pi}
\hat{a}_{\overline{p}}(\omega)^*\frac{\partial\ }{\partial \omega}
\hat{a}_{\overline{p}}(\omega)\,.  \label{posshift}
\end{eqnarray}
The forward-scattering amplitude ${\cal F}$ arises 
in the one-loop diagram without photon emission.  Therefore, the 
second term in Eq.\ (\ref{posshift}) comes from
the quantum correction to the potential rather than
from the reaction to radiation.
(We will discuss this term in the next section.)
Thus, the position shift due to
radiation can be given as
\begin{eqnarray}
\delta z|_{t=0}^{\rm Quantum} & = & - \frac{ie^2}{6\pi pc^3}
\int_{-\infty}^{+\infty}
\frac{d\omega}{2\pi}
\hat{a}_{\overline{p}}(\omega)^*\frac{\partial\ }{\partial \omega}
\hat{a}_{\overline{p}}(\omega)\nonumber \\
& = & \frac{1}{\overline{p}}\int_{-\infty}^{+\infty}dt\,tP_r(t)\,,  
\label{True}
\end{eqnarray}
where $P_r(t)$ given by Eq.\ (\ref{power}) is the power radiated.

One can readily see that Eq.\ (\ref{True}) does 
not agree with Eq.\ (\ref{Shift}) derived in the
Lorentz-Dirac theory.
However, these equations would be identical
if we removed the first term proportional to $\log(v_f/v_i)$
from Eq.\ (\ref{Shift}).  This
term can be traced back to the term proportional to $v\dot{v}$ in 
Eq.\ (\ref{ABRA}) which has arisen due to the substitution
$\ddot{v}v = d(\dot{v}v)/dt - \dot{v}^2$. 
Therefore, Eq.\ (\ref{True}) is reproduced in the classical theory by 
modifying the energy-conservation equation (\ref{LDenergy}) as
\begin{equation}
\frac{d\ }{dt}\left[ \frac{1}{2}mv^2 + V(z)\right]
= -\frac{e^2}{6\pi c^3}\dot{v}^2=-P_r(t)\,.  
\end{equation}
This formula can be interpreted as stating
that the energy of the charged particle is lost
through radiation 
at each time.  In other words, the $\hbar \to 0$ limit of
scalar QED at order $e^2$ can be reproduced by classical 
electrodynamics if momentum conservation is disregarded and
energy conservation is used 
at each moment of time, as far as the position shift is concerned.
This seems reasonable because in the 
quantum theory the energy is conserved 
in the rest frame determined by the potential $V(z)$ but
the momentum is not conserved since the potential
is $z$-dependent. 


\section{Some comments on the forward-scattering amplitude}
\label{forward}

The imaginary part ${\rm Im}\,{\cal F}$ of the forward-scattering 
amplitude is related to the photon emission probability by unitarity,
as we mentioned
before.  The real part ${\rm Re}\,{\cal F}$ represents the
quantum correction to the potential (which may be nonlocal).
To illustrate this fact let us compute the forward-scattering 
amplitude due to a slight change in 
the potential in the form
$V(z)\to V(z) + \delta V(z)$, with $\delta V(z)$ regarded as 
perturbation.  We find 
\begin{eqnarray}
{\cal F} & = & -\frac{2mc^2}{\hbar}\int 
\frac{d^3{\bf P}}{2P_0(2\pi\hbar)^3}
\int d^4 x\, \Phi_{\bf P}^*\delta V(z)\Phi_{\bf p}
\nonumber \\
& = & - \frac{m}{\hbar p}\int_{-\infty}^{+\infty} dz\,\delta V(z)
\end{eqnarray}
to lowest order in $V(z)$ and $\delta V(z)$.
The position shift due to this change in the potential
is 
\begin{equation}
- \hbar \frac{\partial\ }{\partial p}{\rm Re}\,{\cal F}
= -\frac{m}{p^2}\int_{-\infty}^{+\infty} dz\,\delta V(z)\,.
\end{equation}
It can readily be verified that the position shift is given by the 
same formula in the classical theory.
We have not been able to find ${\rm Re}\,{\cal F}$ 
to order $\hbar^0$ in our model but will show that the 
quantum correction to the potential is of order $\hbar^{-1}$.  
This will imply that the electromagnetic
correction to the potential at order $e^2$ cannot be
obtained in the classical theory.

Since the potential is part of the mass term in our model, its quantum 
correction is related to the mass renormalization.
If we let $V(z)=0$ in the mass function given by 
\begin{equation}
[m^2c^2 + 2mV(z)]/\hbar^2 \equiv [m(z)]^2c^2/\hbar^2\,,
\end{equation}
its quantum correction can be computed in the
dimensional regularization with $D=4-2\epsilon$ as
\begin{equation}
\delta m^2 = - \frac{e^2}{16 \pi^2 \hbar c}m^2
\left( \frac{3}{\epsilon} - 3\gamma + 7 - 3 \ln \frac{m^2}{4\pi \mu^2}
\right)\,, \label{msquared}
\end{equation}
where $\mu$ is the renormalization scale and $\gamma$ is Euler's 
constant.  The mass function 
is corrected according to this formula for 
large and positive $z$ for which $V(z)=0$.
For large and negative $z$, the squared mass is
$m^2 + 2mV_{-\infty}/c^2$.  Therefore the quantum correction to the 
mass term is obtained by replacing $m^2$ by this value in 
Eq.\ (\ref{msquared}).  This implies that the difference between 
the quantum corrections to the potential at $z=\pm \infty$ is of order 
$\hbar^{-1}$.  
Hence, this correction cannot be 
obtained in classical electrodynamics\ \footnote{
The corresponding correction to the forward-scattering amplitude is
of order $\hbar^{-2}$.  Since 
the imaginary part of ${\cal F}$
is of order $\hbar^{-1}$, the correction to the potential
does not have an imaginary part at leading order in $\hbar$.}.

If the potential is slowly varying so that the WKB approximation is 
valid, it is reasonable to expect that
the quantum correction to the mass term at leading order in $\hbar$ is 
obtained by 
replacing $m^2$ by $[m(z)]^2$ in Eq.\ (\ref{msquared}) for all $z$.  
Then,
\begin{equation}
\delta [m(z)]^2 
= - \frac{e^2[m(z)]^2}{16\pi^2 \hbar c}
\left( \frac{3}{\epsilon} - 3\gamma + 7 - 3 
\ln \frac{[m(z)]^2}{4\pi \mu^2} \right)\,.
\end{equation}
The quantum correction to the potential can be
expressed by using this formula and Eq.\ (\ref{msquared}) as
\begin{eqnarray}
\delta\left(\frac{V(z)}{mc^2}\right) & = & \frac{1}{2}
\delta \left( \frac{[m(z)]^2}{m^2}\right)\nonumber \\
& = & \frac{3e^2}{32 \pi^2 \hbar c}\frac{[m(z)]^2}{m^2}
\ln \frac{[m(z)]^2}{m^2} 
\label{delV}
\end{eqnarray}
to lowest order in $e^2$.
Since the acceleration is $-V'(z)/m$, this formula shows that the 
correction to the acceleration is ultraviolet finite at leading order
in $\hbar$.


\section{Conclusion}
\label{conclusion}

In this paper we studied a wave packet of a charged scalar particle
moving in the $z$-direction and accelerated by a potential which
depends only on $z$.  Our main conclusion is that, in the limit
$\hbar \to 0$, the position shift
due to the reaction to radiation agrees not with the formula in the
Lorentz-Dirac theory but with that obtained by assuming that the
kinetic energy is lost to radiation according to the 
Larmor formula at each moment of time, at leading order in $e^2$
and in the nonrelativistic approximation.
It will be interesting to study a more general system and see if
the result obtained here can be generalized.

Finally, 
let us discuss the magnitude of the position shift in comparison
with the width of the wave packet.  As we have seen in the 
introduction, the position shift may be much smaller than indicated
by the fact that it is of order $\hbar^0$.
First we note that 
the first term in Eq.\ (\ref{posshif2}), which represents the 
discrepancy between the Lorentz-Dirac theory and the quantum theory, 
is in fact
much smaller than the Compton wavelength of the particle:
\begin{equation}
\frac{e^2}{6\pi mc^3}v_f\log \frac{v_f}{v_i}
= \frac{2\alpha\hbar}{3mc}\left(\frac{v_f}{c}\right)
\log\frac{v_f}{v_i}\ll \frac{\hbar}{mc}\,.
\end{equation}
Thus, it is much smaller than the width of any nonrelativistic
wave packet.  Hence, this term
would be practically unobservable even if it was present.
The second term, which is the position shift in the quantum theory,
grows linearly as a function of the time $|T_0|$ after the 
acceleration has taken place.
For large $|T_0|$ this term is approximately $-E_r |T_0|/p$, 
where $E_r$ is the classical energy of the radiation.  
If the width $\Delta z$ of the wave 
packet is narrowest with $\Delta z\sim \hbar/\Delta p$ 
when the particle is going through acceleration\ \footnote{The wave
packet (\ref{concrete}) is narrowest at $t=0$ when the position is
measured.  One can make this wave packet more general so that 
the case considered here, which is more realistic,
is included by changing the factor $e^{-ipz_0/\hbar}$ to
$e^{-ih(p)/\hbar}$ with $h'(\overline{p}) =z_0$.  
The position shift will be unchanged with this modification.}, 
then after time $|T_0|$ its square behaves like
$(\Delta z)^2 \sim (\hbar/\Delta p)^2 + (\Delta p/m)^2|T_0|^2$.  
The second term dominates
for large enough $|T_0|$.  Therefore, in theory, one can make the 
position shift larger than the width of the wave packet by making 
$|T_0|$ large and letting $\Delta p$ satisfy
\begin{equation}
\Delta p < \frac{m E_r}{p} \sim 
\frac{\alpha \hbar m a^2 t_a}{p c^2}\,,
\end{equation}
where $a$ is the typical acceleration and where $t_a$ is the
duration of the acceleration\ \footnote{Since the second term in
Eq.\ (\ref{posshif2}) is the lowest-order term in $V(z)$ that grows
linearly in $|T_0|$, the fact that it is much larger than the 
first term, which is of lower order in $V(z)$, does not imply that our
approximation breaks down since the latter is 
independent of $|T_0|$.}.
Then, with $p=mv$,
\begin{equation}
\Delta z \sim \frac{\hbar}{\Delta p} 
\agt \frac{1}{\alpha}\left( \frac{c}{a t_a}\right)^2 z_a \gg z_a\,,
\end{equation}
where $z_a = v t_a$ is the length of the interval in $z$ where the
acceleration occurs.  Thus, the wave packet would be
much wider than the region of acceleration if the position shift were
to be larger, in theory, than the width of the wave packet at a
later time.

\section{Erratum}
\label{correction}

In finding the $\hbar\to 0$ limit of the scattering amplitude 
[Eq.\ (\ref{intermed})] we
erroneously discarded surface terms.  We correct this error here.
This correction changes our conclusion completely: the
position shift in the Lorentz-Dirac theory agrees with that in
quantum field theory in the $\hbar\to 0$ limit.

Since the function $K(z)$ in Eq.\ (\ref{gkz}) can be approximated
in the limit $\hbar \to 0$ 
by $\omega/v_p(z)$ [see Eq.\ (\ref{Kappr2})] we have
\begin{eqnarray}
G(k_z,P,p) & = & 2p\int_{-\infty}^{+\infty} dz \exp
\left( i\int_0^z \omega/v_p(z')\,dz'\right)\nonumber \\
& = & 2p\int_{-\infty}^{+\infty} v_p(t) e^{i\omega t}\,.
\end{eqnarray}
(Note that $P\to p$ in the $\hbar \to 0$ limit.)
Since this integral is ill-defined, we insert a damping factor
$\chi(t)$ which takes the value 1 while $a_p(t)= v_p'(t)\neq 0$ and
goes to zero smoothly as $t\to \pm \infty$.  Thus,
\begin{equation}
G(k_z,P,p)= 2p\int_{-\infty}^{+\infty} v_p(t)\chi(t)e^{i\omega t}\,.
\end{equation}
The quantity $\hat{a}_p(\omega)$ in Eq.\ (\ref{intmed}) needs to
be replaced by
$$
-i\omega \int_{-\infty}^{+\infty}v_p(t)\chi(t)e^{i\omega t}
= \hat{a}_p(\omega) + \int_{-\infty}^{+\infty}v_p(t)\chi'(t)
e^{i\omega t}\,.
$$
Then the contribution of radiation reaction to the position shift is
\begin{eqnarray}
\langle z\rangle_{t=0}^{\rm rad}
& = & \frac{e^2}{6\pi c^3}\int_{-\infty}^{+\infty}\frac{d\omega}{2\pi}
\nonumber \\
&& \times 
\left[\hat{a}_p(-\omega)+\int_{-\infty}^{+\infty}v_p(t')\chi'(t')
e^{-i\omega t'}dt'\right] \nonumber \\
&& \times \int_{-\infty}^{+\infty}dt\frac{\partial v_p(t)}{\partial p}
\chi(t)e^{i\omega t}\nonumber \\
& = & \frac{e^2}{6\pi c^3}\int_{-\infty}^{+\infty} dt \nonumber \\
&& \times
\left\{ a_p(t)\frac{\partial v_p(t)}{\partial p}
+ \frac{1}{2}v_p(t)\frac{\partial v_p(t)}{\partial
p}\frac{d\ }{dt}\left[\chi(t)\right]^2\right\}\,. \nonumber \\
\label{crit}
\end{eqnarray}

The quantity $\partial v_p(t)/\partial p$ can be found as follows.
Note that
\begin{equation}
t = \int_{z_0}^{z}\frac{m}{\sqrt{p^2-2mV(\xi)}}d\xi
\end{equation}
and hence
\begin{equation}
dt = \frac{1}{v_p(t)}dz_p - \frac{v_f}{m}\left(\int_0^t \frac{d\tau}
{\left[v_p(\tau)\right]^2}\right) dp\,.
\end{equation}
This can be re-arranged as
\begin{equation}
dz_p = v_p(t)dt + \frac{v_f}{m}v_p(t)\left(\int_0^t \frac{d\tau}
{\left[ v_p(\tau)\right]^2}\right)dp\,.
\end{equation}
By using equality of mixed partial derivatives,
\begin{equation}
\frac{\partial\ }{\partial p}\left(\frac{\partial z_p}{\partial t}
\right) = \frac{\partial\ }{\partial t}\left(\frac{\partial z_p}
{\partial p}\right)\,,
\end{equation}
we obtain
\begin{equation}
\frac{\partial\ }{\partial p}v_p(t)
= \frac{v_f}{m}\left[
\frac{1}{v_p(t)} + a_p(t) \int_0^t \frac{d\tau}{\left[v_p(\tau)\right]^2}\right]\,.
\end{equation}
By substituting this in Eq.\ (\ref{crit}) we find
\begin{eqnarray}
\langle z\rangle_{t=0}^{\rm rad}
& = & \frac{e^2 v_f}{6\pi mc^3}\int_{-\infty}^{+\infty} dt \nonumber \\
&& \times
\left\{ \frac{a_p(t)}{v_p(t)} + \left[a_p(t)\right]^2 \int_0^t
\frac{d\tau}
{\left[v_p(\tau)\right]^2} \right. \nonumber \\
&& \left. - \frac{1}{2}\frac{d\ }{dt}\left[
a_p(t)\int_0^t\frac{d\tau}{\left[v_p(\tau)\right]^2}\right]
\right\}\nonumber \\
& = & \frac{e^2 v_f}{6\pi mc^3}
\left\{ \log \frac{v_f}{v_i} \right.\nonumber \\
&& \left. - \int_{-\infty}^0 \frac{d\tau}{\left[v_p(\tau)\right]^2} 
\left(
\int_{-\infty}^\tau \left[a_p(t)\right]^2 dt\right)\right\}\,,\nonumber \\
\end{eqnarray}
where we have used the assumption that $a_p(t) = 0$ if $t \geq 0$.
This is exactly the formula obtained by integrating Eq.\ (\ref{preDelz})
found using the Lorentz-Dirac theory
from $t=-\infty$ to $t=0$.  Notice that we did not
use the assumption that $2m|V(z)| \ll p^2$.


\acknowledgments

The author thanks Bernard Julia and Mitch Pfenning for useful 
discussions and the anonymous referee for helpful comments.

\

\appendix

 
\section{Calculations without choosing physical polarization
states}

If we do not use physical polarization states for photons
and stick to the Fock space
with indefinite metric for the Feynman gauge, the one-photon final 
state is not given by Eq.\ (\ref{final}) but by
\begin{eqnarray}
|f,{\rm 1\gamma}\rangle & = & 
i\int\frac{d^3{\bf k}}{2\hbar \omega(2\pi)^3}
\int\frac{d^3{\bf P}}{2P_0(2\pi\hbar)^3}
\int\frac{d^3{\bf p}}{\sqrt{2p_0}(2\pi\hbar)^3} \nonumber \\
&& \times f({\bf p})
{\cal A}^\mu({\bf k};{\bf P};{\bf p})
b_\mu^\dagger({\bf k}) A^\dagger({\bf P})|0\rangle\,,
\end{eqnarray}
where
\begin{equation}
{\cal A}_\mu ({\bf k};{\bf P};{\bf p})
 = - ie\hbar c^2
\int dt\,d^3{\bf x} 
\,\Phi_{\bf P}(t,{\bf x})^*
\stackrel{\leftrightarrow}{\partial}_\mu
\Phi_{\bf p}(t,{\bf x})\,.
\end{equation}
One may assume that
${\bf P}_\perp$ and ${\bf p}_\perp$ are negligibly small as before.
Let us define $G_\mu(k_z,P,p)$ by
\begin{eqnarray}
&& {\cal A}_\mu ({\bf k},{\bf P},{\bf p})
\nonumber \\
&& \equiv ec^2 G_\mu (k_z,P,p)\nonumber \\
&& \ \ \ \ \times 
(2\pi\hbar)^3 \delta^2({\bf P}_\perp
+\hbar {\bf k}_\perp -{\bf p}_\perp)
\delta(P_0+\hbar \omega -p_0)\,.\nonumber \\
\end{eqnarray}
It is clear that $G_x$ and $G_y$ are negligibly small and that $G_0$ and
$G_z$ are independent of the direction of ${\bf k}_\perp$.  We have
\begin{equation}
G_z(k_z,P,p) = G(k_z,P,p)\,, 
\end{equation}
where $G(k_z,P,p)$ is defined by Eq.\ (\ref{Gkz}). 
The $0$-component $G_0(k_z,P,p)$ is
\begin{eqnarray}
G_0(k_z,P,p)
& = & -\sqrt{Pp}\,\frac{p_0+P_0}{c}
\int \frac{dz}{\left[\kappa_P(z)\kappa_p(z)\right]^{1/2}}\nonumber \\
&& \times \exp\left[ i
\int_0^z K(z')\,dz'\right]\,,
\end{eqnarray}
where $K(z)$ is defined by Eq.\ (\ref{defK}).
Using a formula similar to Eq.\ (\ref{gzform}),
we arrive at
\begin{eqnarray}
G_0(k_z,P,p) & = & -i\sqrt{Pp}\,\frac{p_0+P_0}{c}
\int dz\,\nonumber \\
&& \times \frac{d\ }{dz}
\left\{\left[\kappa_P(z)\kappa_p(z)\right]^{1/2}K(z)\right\}^{-1}
\nonumber \\
&& \times \exp\left[ i \int_0^z K(z')\,dz'\right]\,.
\end{eqnarray}
At lowest order in $\hbar$ we can let $P=p$, $P_0=p_0$,
$\kappa_P(z)=\kappa_p(z) \equiv mv_p(z)$ and use the
approximation (\ref{Kappr}).
Thus, we find
\begin{eqnarray}
G_0(k_z,P,p) & = & 
-\frac{2ipp_0}{c}\int_{-\infty}^{+\infty}dz
\frac{k_z v_p'(z)}{m\left[\omega-v_p(z)k_z\right]^2}
\nonumber \\
&& \times 
\exp\left\{ i\int_0^z \left[\omega/v_p(z')-k_z\right]\,dz'\right\}\,.
\nonumber \\
\end{eqnarray} 
Then, by letting $p_0 = mc^2$ and $k_z = 0$ on the right-hand side 
[because $\omega \gg v_p(z)k_z$] and comparing the result with 
Eq.\ (\ref{Gappr}) we find
\begin{equation}
G_0(k_z,P,p) = -\frac{ck_z}{\omega}G(k_z,P,p)\,.
\end{equation}
Thus, the only difference between the calculations
here and those in Secs.~\ref{amplitude} and \ref{position} is that the
factor $|\epsilon^{(1)z}({\bf k})G(k_z,P,p)|^2$ is replaced here by
$-G_\mu(k_z,P,p)G^\mu(k_z,P,p)=(1-k_z^2 c^2/\omega^2)|G(k_z,P,p)|^2$
in the ${\bf k}$-integral for the final results.  For a given $\omega$
the average of $1-k_z^2 c^2/\omega^2$ over the solid angle is $2/3$.
Since the average of $|\epsilon^{(1)z}({\bf k})|^2$ is also $2/3$, we
arrive at the same transition probability (\ref{transprob}) 
and position shift (\ref{True}).

 
\section{A derivation of the position expectation value}

In this Appendix we derive Eq.\ (\ref{Peval}) which was used to find 
the expected value of the $z$-coordinate of the particle.
A systematic method to obtain a wave function for a one-particle state
in the scalar field theory has been given by Feshbach and 
Villars~\cite{FV,Dav}.  We will not use this method explicitly, but
the derivation we describe here is based on the idea spelled out in 
Ref.\ \cite{FV}.

Recall that the charge density operator is
\begin{equation}
\rho = \frac{i}{\hbar c^2}\left(\psi^\dagger \partial_t \psi
- \psi \partial_t \psi^\dagger\right)\,.
\end{equation}
Let us define 
\begin{equation}
\rho_i(t, {\bf x}) \equiv \langle i|\rho(t,{\bf x})|i\rangle\,,
\end{equation}
where the state $|i\rangle$ is defined by Eq.\ (\ref{wave_packet}).  
Since this state is a one-particle state, the charge density 
$\rho_i(t,{\bf x})$ coincides with
the probability density function for the position of the particle at 
each time up to a normalization factor (which will turn out to be one).
By a straightforward calculation we find
\begin{equation}
\rho_i (t,{\bf x}) = i\hbar
\phi_i(t,{\bf x})^*\stackrel{\leftrightarrow}{\partial_t}
\phi_i (t,{\bf x})\,,  \label{rhoi}
\end{equation}
where
\begin{eqnarray}
\phi_i(t,{\bf x}) & \equiv & \frac{1}{\hbar c}
\langle 0|\psi(t,{\bf x})|i\rangle\nonumber \\
& = & \int \frac{d^3{\bf p}}{\sqrt{2p_0}(2\pi\hbar)^3}f({\bf p})
\Phi_{\bf p}(t,{\bf x})\,. \label{Phiapprox}
\end{eqnarray}
Then it follows from orthonormality (\ref{orthon}) of 
$\Phi_{\bf p}(t,{\bf x})$ and the normalization condition
(\ref{normall}) for $f({\bf p})$ that
\begin{equation}
\int d^3{\bf x}\,\rho_i(t,{\bf x}) = 1\,.
\end{equation}
Thus, the function $\rho_i(t,{\bf x})$ is the normalized probability
density function for the position of the particle.
By substituting Eq.\ (\ref{Phiapprox}) in Eq.\ (\ref{rhoi}) we obtain
\begin{eqnarray}
\rho_i(t,{\bf x}) & = &
\frac{1}{2}\int \frac{d^3{\bf p}}{(2\pi\hbar)^3}
\int \frac{d^3{\bf p}'}{(2\pi\hbar)^3}
\left( \sqrt{\frac{p'_0}{p_0}} + \sqrt{\frac{p_0}{p'_0}}\right)
\nonumber \\
&& \ \ \times 
f({\bf p}')^*f({\bf p})\Phi_{{\bf p}'}(t,{\bf x})^*
\Phi_{\bf p}(t,{\bf x})\,, \label{rhoi2}
\end{eqnarray}
where $p_0$ is defined by Eq.\ (\ref{energy}) with $p'_0$ 
similarly defined. 

The expected value of the $z$-coordinate at $t=0$ is
\begin{equation}
\langle z\rangle_{t=0}^{\rm off} = \int d^3{\bf x}\, z\,\rho_i(0,{\bf
x})\,.  \label{rhoi3}
\end{equation}
By using the assumption that the wave packet is located far into
the region where $V(z)=0$ at $t=0$
we may approximate the mode function $\Phi_{\bf p}(0,{\bf x})$ at 
$t=0$ in Eq.\ (\ref{rhoi2}) as
\begin{equation}
\Phi_{\bf p}(0,{\bf x}) \approx \exp\left[\frac{i}{\hbar}\left(
pz + {\bf p}_\perp\cdot {\bf x}_\perp\right)\right]\,.
\end{equation} 
With this approximation we can write
\begin{equation}
z \Phi_{\bf p}(0,{\bf x}) \approx
-i\hbar \frac{\partial\ }{\partial p} \Phi_{\bf p}(0,{\bf x})\,.
\end{equation}
By substituting this
formula in Eq.\ (\ref{rhoi3}) and integrating by parts we find
\begin{widetext}
\begin{eqnarray}
\langle z\rangle_{t=0}^{\rm off}  
& = & \frac{i\hbar}{2} \int \frac{d^3{\bf p}}{(2\pi\hbar)^3}
\int \frac{d^3{\bf p}'}{(2\pi\hbar)^3}\,\frac{\partial\ }{\partial p}
\left[ \left( \sqrt{\frac{p_0}{p'_0}} + \sqrt{\frac{p'_0}{p_0}}\right)
f({\bf p}')^*f({\bf p})\right]
\int d^3{\bf x} \Phi_{{\bf p}'}(0,{\bf x})^*
\Phi_{\bf p}(0,{\bf x})\nonumber \\
& = & 
i\hbar \int \frac{d^3{\bf p}}{(2\pi\hbar)^3}
\int \frac{d^3{\bf p}'}{(2\pi\hbar)^3}\nonumber \\
&& \times
\left[ \frac{c^2}{2}\left( \frac{1}{\sqrt{p'_0p_0}} 
- \sqrt{\frac{p'_0}{p_0^3}} \right)f({\bf p}')^*f({\bf p}) + 
f({\bf p}')^*\frac{\partial\ }{\partial p}f({\bf p})\right]
(2\pi\hbar)^3\delta^3({\bf p}-{\bf p}') \nonumber \\
& = & 
i\hbar \int \frac{d^3{\bf p}}{(2\pi\hbar)^3}
f({\bf p})^*\frac{\partial\ }{\partial p}f({\bf p})\,.
\end{eqnarray}
We immediately obtain Eq.\ (\ref{Peval}) from this formula by 
integration by parts.
\end{widetext}


\end{document}